# "Water Buckyball" Terahertz Vibrations in Physics, Chemistry, Biology, and Cosmology

Keith Johnson

*Massachusetts Institute of Technology, P.O. Box 380792, Cambridge, MA 02238-0792 and
HydroElectron Ventures Inc., 1303 Greene Avenue Suite 102, Westmount, QC, Canada, H3Z 2A7*

**Abstract**

**Pentagonal dodecahedral water clusters – "water buckyballs" - and arrays thereof are shown from first-principles electronic-structure calculations to possess unique terahertz-frequency vibrational modes in the 1-6 THz range, corresponding to O–O–O "bending", "squashing", and "twisting" "surface" distortions of the clusters. The cluster LUMOs are huge "Rydberg" "S"-, "P"-, "D"-, and "F"- like molecular orbitals that accept an extra electron via optical excitation, ionization, or electron-donation from interacting atoms or molecules. Dynamic Jahn-Teller coupling of these "hydrated-electron" orbitals to the THz vibrations promotes such water clusters as oxygenates for destroying particulate matter in hydrocarbon fuel combustion, as candidates for reducing the large contribution of water vapor to atmospheric global warming, as the focus for harvesting methane from water clathrates, and as vibronically active "structured water" essential to biomolecular function such a*s* protein folding. In biological "microtubules", confined water-cluster THz vibrations may induce their "quantum coherence", communicated by Jahn-Teller phonons via coupling of the THz electromagnetic field to the water clusters' large electric dipole moments. Also discussed are the roles of networked water buckyball clathrates in the structure and possible high-$T_c$ superconductivity of hydrated-electron supercooled water. Lastly, astrophysical water buckyballs may constitute thermalizing "cosmic dust", contribute to the cosmic background radiation, be a candidate for "Rydberg baryonic dark matter", and absorb the virtual photons of vacuum fluctuations above their cutoff THz vibrational frequency, leaving only the low-frequency virtual photons to be gravitationally active, thus possibly explaining the small magnitude of dark energy.**

**Introduction**

Scientific interest in water nanoclusters has been motivated by their possible roles in biology [1,2] atmospheric science [3,4], and astrophysics [5], as well as by their relevance to the structure and properties of liquid water [6]. In a recent letter by this author and collaborators [7], it has been suggested that because of their large electric dipole moments, protonated water clusters such as $(H_2O)_{21}H^+$ shown in Fig. 1 may contribute significantly to the observed strong THz emission from water vapor under optical stimulation. Experiment and theory agree that not only can such water clusters be produced, but they also occur optimally in certain numbers (so-called "magic numbers") and configurations of water molecules [3,8-10]. Prominent among the magic-number water clusters that have been identified are ones having the "buckyball-like" pentagonal dodecahedral structure [9,10] shown in Fig. 2a. These clusters have a closed, ideally icosahedral symmetry formed by 20 hydrogen-bonded water molecules, with their oxygen atoms at the vertices of 12 concatenated pentagons and with 10 free exterior hydrogen atoms, as illustrated for $(H_2O)_{21}H^+$ and $(H_2O)_{20}$ clusters in Figs. 1d and 3a., respectively.

**Electronic Structure and THz Vibrations**

Fig. 1 shows the computed density-functional ground-state molecular-orbital energies and vibrational modes of the $(H_2O)_{21}H^+$ or $(H_2O)_{20}H_3O^+$ cluster, which from molecular-dynamics simulations are qualitatively unchanged at temperatures well above 100 degC, where the cluster remains remarkably intact. Of particular importance are the "squashing" and "twisting" vibrational modes of the otherwise ideal pentagonal dodecahedral cluster shown in Fig. 2a. Density-functional calculations for the electrically neutral pentagonal dodecahedral water cluster, $(H_2O)_{20}$, and arrays thereof shown in Fig. 3 have also been performed, yielding the vibrational modes displayed in Fig. 4. These results are very similar to those for $(H_2O)_{21}H^+$ in Fig. 1. Common to the water clusters studied are: (1) lowest unoccupied (LUMO) energy levels like those shown in Fig. 1a, which correspond to the huge "Rydberg" "S"-, "P"-, "D"- and "F"-like cluster "surface" orbital wavefunctions shown in Fig. 1b; and (2) bands of vibrational modes between 1 and

6 THz, shown in Figs 1 and 4, due to O-O-O "squashing" (or "bending") and "twisting" motions of the type illustrated in Fig 2a between adjacent hydrogen bonds. The vectors in Figs. 1 and 4 represent the directions and relative amplitudes for the lowest THz modes corresponding to the O-O-O "bending" (or "squashing") motions of the water-cluster "surface" oxygen ions. Surface O-O-O bending vibrations of water clusters in this energy range have indeed been observed experimentally [11]. Near-ultraviolet excitation of an electron from the HOMO to LUMO (Fig 1a) can put the electron into the bound "S"-like cluster molecular orbital mapped in Fig 1b. Occupation of the LUMO produces a metastable bound state of such a water cluster, even when an extra electron is added, the so-called "hydrated electron". In contrast, a water monomer or dimer has virtually no electron affinity. Infrared absorption can further excite the cluster LUMO "S"- like electron into a higher LUMO such as the "P"-like cluster orbital in Fig. 1b. Infrared excitations within the LUMO manifold can then decay vibrationally according to the Franck-Condon principle. The lowest-frequency vibrations of water clusters induced by optical pumping should be the 1-6 THz "surface" modes of Figs. 1c,d and 4.

**Chemical Reactivity and Energy**

While liquid water is not usually thought of as an "active" substance, hydrated-electron water buckyballs are another matter due to the effectively large reactive cross sections of the cluster "surface" molecular orbitals mapped in Fig. 1b. Experimental studies of the thermal energy chemical reactions of size-selected hydrated-electron clusters $(H_2O)_n^{-1}$ attribute their strong reactivity to the spatial delocalization of the excess electron at the cluster surface [12]. Thus charged water buckyballs can function as electron reservoirs for chemical reactions involving electron transfer to or from the reacting species. In the oxidation of carbon compounds, hydrated-electron delocalized orbitals originating on the water-cluster surface oxygen atoms can readily overlap with the reactive fuel carbon (*e.g.* p$\pi$) orbitals, promoting oxidation. The proximity of the highest occupied dodecahedral water-cluster excess-electron "S" orbital to the lowest unoccupied, nearly degenerate cluster "$P_x,P_y,P_z$" orbitals (Fig. 1a,b) suggests the strong possibility of coupling between the hydrated electron and the THz-frequency cluster "squashing" modes shown in Figs. 1, 2a, and 4 via the pseudo or dynamic Jahn-Teller (JT) effect [13] Such cluster "squashing" modes associated with icosahedral symmetry are known to be JT-active in the case of electron-doped buckminsterfullerene [14]. JT coupling in the water clusters leads to a prescribed symmetry breaking of the pentagonal dodecahedron along the THz-frequency vibrational squashing mode coordinates $Q_s$, lowering the cluster potential energy from A to the equivalent minima A' shown in Fig. 2b. Because of the large JT-induced vibrational displacements (large $Q_s$) of water-cluster surface oxygen atoms, the energy barrier for expulsion of water oxygen or OH radicals and their oxidative addition to reactive carbon atoms is lowered from $E_{barrier}$ to $E'_{barrier}$, as shown in Fig. 2b. A more detailed discussion of the JT effect as an activation mechanism in chemical reactions and catalysis is found in ref. [13]. The practical use of water clusters stabilized in nanoemulsions to break down and more completely burn particulate matter (smoke/soot) precursor molecules, as well as serving as a combustion oxygenate in fossil fuels, has been demonstrated [15,16]. The vibrational coupling of water buckyball THz "squashing" mode to a THZ "bending" mode of an anthracene soot precursor molecule, the first step in its decomposition, is shown in Fig. 5c,d. Water clusters could likewise be used to catalyze the efficient breakdown of switchgrass cellulose for biofuel synthesis. Finally, water-cluster THz modes may facilitate electron and proton transfer in photosynthesis [17].

**The Greenhouse Effect and Global Warming**

Contrary to popular belief, the global greenhouse effect may have more to do with atmospheric water than gases such as carbon dioxide and methane [18]. Water vapor causes 36-70% of the greenhouse effect on Earth, not including clouds, while $CO_2$ causes only 9-26%, and $CH_4$, only 4-9%. Although increases of $CO_2$ are indeed a source of the enhanced greenhouse effect, and thus global warming, the contribution of atmospheric water is rarely discussed because, unlike most other gases, the distribution of atmospheric water varies strongly with altitude, terrestrial location, and time, while water vapor changes to the liquid and solid phases at terrestrial temperatures. Water vapor is conventionally viewed as a gas of individual $H_2O$ molecules. However, from spectroscopic studies [3,4], atmospheric water vapor is known to be a natural source of clusters of water molecules, including protonated water clusters such as $(H_2O)_{21}H^+$. A recent quantum chemistry study of smaller water clusters, such as trimers, tetramers, and pentamers, has concluded that they too may contribute significantly to global warming [19].

Water clusters like those shown in Figs 1 and 4 store more heat than do separate water molecules because of the many more cluster vibrational degrees of freedom, such as shown by their unique cluster "surface" vibration vectors. The heat storage capacity of water buckyballs approaches that of bulk water. Thus it is not surprising that even a modest population of such water clusters in the troposphere can contribute significantly to the greenhouse effect and may help explain why water vapor is so much more potent as a greenhouse gas than carbon dioxide alone. Even slight increases of ocean temperature produce significant increases in the evaporation of water molecules and clusters thereof into the atmosphere. Combined with man-made sources of water vapor, such as from industry, terrestrial vehicles, ships, and aircraft, increases of atmospheric water vapor and therefore water clusters due to evaporation from seas, lakes, and land seem to be insurmountable. Even the highly touted fuel cell produces water vapor as a principal product.

As shown in Figs. 1 and 4, water buckyballs and arrays thereof have the unique property that their vibrational frequencies extend into the THz region of the electromagnetic spectrum, whereas the THz spectra of $H_2O$ and $CO_2$ monomers are associated with pure rotations or rovibrations. Atmospheric water vapor is known to be a strong absorber of THz radiation. Through a collaboration with the R.P.I. Center for Terahertz Research, we have shown that optically pumped ambient water vapor is also a strong emitter of THz radiation, possibly associated with water cluster THz vibrational modes [7].

The calculated ground-state dipole moment of the $(H_2O)_{21}H^+$ cluster is approximately 10 Debyes, as compared with the 1.86 D moment of water monomer. THz vibrations induced by optical pumping produce large oscillating electric dipole moments exceeding 10D. Since the power radiated from an oscillating dipole is proportional to the square of the dipole moment, an optically excited water cluster ion like $(H_2O)_nH^+$ is a potentially strong source of THz radiation. Ambient water vapor containing $10^{17}$ cm$^{-3}$ water monomers typically contains only $10^4$ to $10^6$ $(H_2O)_nH^+$ clusters cm$^{-3}$, whereas neutral water dimer and cluster concentrations are as high as $10^{10}$ to $10^{12}$ cm$^{-3}$ [3,4]. Nevertheless, isolated $(H_2O)_nH^+$ clusters can in theory be pumped collectively by photons into resonant vibrational (phonon) modes of lowest frequencies 1-6 THz, analogous to a Bose-Einstein condensation, acquiring giant electric dipoles of the order of $10^5$ to $10^7$ D cm$^{-3}$ [20]. This greatly enhances the possible contribution of water clusters to radiation absorption and emission and has been argued to explain the strong far-infrared and submillimeter absorption of radiation in Earth's atmosphere [3,4].

Water buckyballs and arrays thereof can hold an extra electron in their LUMOs – the "hydrated electron". However, the addition of more electrons to water clusters will break them down into hydrogen and oxygen, or at least convert some of their infrared radiation to harmless THz emission. In principle therefore, the injection of electrons into the troposphere should reduce the contribution of water molecule clustering to atmospheric heat storage and global warming. This could be accomplished safely from the many commercial aircraft flying daily through the troposphere by exposing photoelectron-emitting panels that would release electrons under solar radiation once the plane was at altitude. Although ozone is likely to be a byproduct, tropospheric ozone is beneficial for absorbing dangerous ultraviolet radiation.

**Methane Hydrate Clathrates and Energy**

Methane hydrate is a combination of methane gas and water formed into a substance that looks like ice but it is unstable at standard temperature and pressure [21]. The structure is metastable due to weak bonding between the gas and the water molecules and hydrogen-bonding between water molecules within the dodecahedral clathrate cage structure shown in Fig. 5a. Methane hydrates are found naturally under the ocean floor and in permafrost. There is concern that global warming could melt some shallow methane deposits, releasing millions of tons of this potent greenhouse gas into the air.

Worldwide amounts of carbon bound in methane and other gas hydrates are conservatively estimated to total twice the amount of carbon to be found in all other known fossil fuels on Earth. Safe harvesting of methane from hydrates could provide an enormous energy resource. If terahertz radiation in the 1–6 THz range were to be applied to methane hydrate to excite the large-amplitude gas-hydrate vibrations of the type shown in Fig.5b, the vibrations should close the clathrate HOMO-LUMO energy gap (Fig. 1a), pouring electrons from the bonding into the antibonding methane-hydrate molecular orbitals and thereby causing the controlled release of methane gas from the water-clathrate cages.

**Protein Folding and Drug Receptors**

Pentagonal water clusters have been experimentally identified as key to the hydration and stabilization of biomolecules [1,2]. Such examples indicate the tendency of water pentagons to form closed geometrical structures at biomolecular interfaces including amino acids (Fig. 5e) and nucleotides (DNA, RNA) [22,23]. The "restructuring" or clustering of water molecules may even determine biological cell architecture [24]. Approximately 70 percent of the human body is water by weight. Much of that water is believed to be restructured or nanostructured, which affects biomolecular processes including protein stability, enzyme activity, and proton transfer [25,26].

Some diseases (*e.g.* Alzheimer's, Parkinson's, type II diabetes, and cataracts) are associated with the "misfolding" of proteins. Water - "restructured" as nanoclusters of the above-described type – plays a key role in the proper folding of proteins [27]. The misfolding of proteins, making them dysfunctional and disease-causing, is likely associated with the failure of water molecules to congregate in clusters that properly interact with the protein. The development of drugs to treat such diseases should therefore be focused on the restoration of water clustering at the protein interfaces [28]. Water cluster "surface" THz vibrational modes like the ones shown above are especially important because they couple or "resonate" with THz-frequency "bending" vibrations of the amino-acid residues in proteins. An example is shown in Fig. 5e,f for the protein, insulin, where it has been established experimentally that a "hemispherical" pentagonal dodecahedral water cluster "clathrates" a "valyl" amino acid residue [27]. This property is also key to optimizing the delivery of drugs to and their interaction with drug-receptor sites, where THz vibrations of water clusters clathrating the drug may provide the resonance needed to restore the interfacial water restructuring essential for proper protein folding. Formation of hydrogen bonds between a drug molecule and a water molecule will polarize the latter, resulting in further hydrogen bonding to other water molecules and the formation of water clusters that play a key role in drug receptor identification. Even in the absence of hydrogen bonding, drug molecules will tend to cause water restructuring and the formation of cage-like structures, which implies that the drug molecule will have a "water signature". Finally, the application of infrared radiation to restructure water into clathrate structures [29] near cancerous (*e.g.* skin or breast) tissue, which is known to harbor more "liquid-like" water [30], could possibly complement drug treatment.

**Quantum Coherence in Biology**

There are environments in biology where water molecules may be confined in linear or "filamentary" arrays like the one shown in Fig. 6a instead of in globular arrays like those shown in Fig. 3b,c. One example is the "microtubule", a principal component of the cytoskeleton - hollow polymeric cylinders approximately 25 nm in diameter composed of the protein, 'tubulin". Microtubules are dynamic structures that help to determine biological cell shape and facilitate intracellular transport [31]. As a first step toward modelling the water confined to a microtubule, a linear array of pentagonal dodecahedral water clusters has been confined to a cell with periodic boundary conditions on the ends of the cell and an electron added to the array to simulate electron donation from a surrounding microtubule. Density-functional calculations of the lowest energy structure were performed and produce the stable configuration and vibrational spectrum shown in Fig. 6. Two primary bands of THz vibrational modes are found, in common with the results obtained for globular arrays of water buckyballs shown in Fig. 4. The lowest-frequency THz vibrational modes, such as the one displayed in Fig, 6c, communicated by Jahn-Teller phonons [13] via coupling of the THz electromagnetic field to the water clusters' extraordinarily large electric dipole moments, may lead to the "quantum coherence" or "Bose-Einstein-like" condensation of interacting water clusters confined to microtubules [32]. While the human body is 70 percent water by weight, the normal human brain is nearly 90 percent water. It is possible that this quantum-coherent dynamical Jahn-Teller system of water clusters and the quantized THz electromagnetic field confined within the hollow inner cores of brain microtubules may be relevant to conscious thought processes, consistent with controversial ideas promoted by Penrose and Hameroff [33].

**Supercooled Water and High-$T_c$ Superconductivity**

Recent X-ray diffraction studies of supercooled water have revealed the presence of pentagonal dodecahdral water-cluster clathrate structures similar to those shown in Fig. 3 [34]. The clathrate structures become dominant with decreasing temperature. The addition of electrons to supercooled water should occupy LUMO molecular orbitals of the type shown in Fig. 1b. According to a molecular-orbital criterion for high-$T_c$ superconductivity [35], the occupation of LUMO p$\pi$ orbitals shown in Fig. 1b, coupled to the THz vibrations (phonons) via the dynamic Jahn-Teller effect [13], should result in superconductivity at a temperature of approximately -43 degC. This is coincidentally very close to the lowest temperature of -42 degC actually achieved for supercooled water, which at that temperature forms a glassy substance with rapid cooling [34]. According to ref. 35 the local phonons responsible for Cooper pairing of the added electrons and their superconductivity have a cut-off frequency of 32 THz, which is exactly the cut-off frequency of the water-cluster THz vibrations shown in Figs, 1 and 4. This is a prediction that needs to be confirmed by experiment. If confirmed, negatively charged supercooled glassy water at -43 degC would be the highest-$T_c$ superconductor observed to date.

**Baryonic Dark Matter and Dark Energy**

It was originally suggested by Layzer and Hively [36] that the spectrum and isotropy of the cosmic microwave background might be attributable to thermalization by "cosmic dust" in the form of hollow, spherical shells of high dielectric constant. Using a formula for the absorption-plus-scattering cross-section of dielectric spheroids [37], it was argued that a relatively low density of dust could thermalize the radiation produced by collapsing objects of galactic mass at a redshift z = 10. Water clusters have been argued to occur in deep space [5]. Because of their extraordinarily high electric dipole moment, approximately spherical "shells" of O-H bonds (Figs. 1 and 3), and strong absorption plus scattering of THz radiation, optically-pumped water clusters satisfy the conditions proposed in ref. [36]. Recent astronomical observations [38] push back the epoch of protogalaxy formation and reionization to a redshift z = 10, *i.e.* to a time just 460 million years after the Big Bang. At z = 10, the distinctive 1-6 THz vibrational manifolds of the protonated cluster $(H_2O)_{21}H^+$ (Fig. 1), of "globular" $(H_2O)_n$ clusters (Fig 4), and of "filamentary" $(H_2O)_n^-$ clusters (Fig. 6) are red-shifted to the cosmic microwave (GHz) background spectrum, suggesting water clusters could constitute thermalizing "cosmic dust" of the type suggested in ref. [36]. It is an open question whether water molecules and clusters thereof actually existed at z = 10. Nevertheless, water masers, typically found in dense molecular clouds associated with supermassive black holes at the centers of nearby galaxies, but not expected in more distant galaxies, have recently been observed at surprisingly high densities and luminosities in a high-redshift (z = 2.64) galaxy more than 11 billion light-years away when the Universe was only one sixth the age it is now [39]. In other words, water molecules and likely clusters thereof may have been common in the early universe. Hydrogen and oxygen, respectively, are the first and third most abundant elements in the Universe, so one would expect water to be plentiful. In interstellar clouds, stable, protonated water clusters can easily be created from ice-covered cosmic grains by dissociative ionization and cosmic-ray bombardment [5]. Because their highly delocalized "S"-, "P"-, "D"- and "F"-like LUMOs (Fig. 1a,b), even with an added electron, don't absorb or emit visible light, water buckyballs may therefore be viewed as a form of "Rydberg baryonic dark matter" [40].

Beck and Mackey [41] suggest that if the "dark energy" responsible for the accelerated expansion of the Universe is equated to the cosmic vacuum energy, then gravitationally active vacuum fluctuations must have a cutoff frequency of the order of 1.7 THz. In other words, the virtual photons associated with vacuum fluctuations should be gravitationally active only below this frequency to be consistent with the magnitude of dark energy. Perhaps fortuitously, 1.7 THz is the same order of magnitude as the cutoff frequencies of typical water clusters shown in Figs. 1, 4 and 6. Although water clusters have a low cosmic density compared to atomic hydrogen, they satisfy ref. [36] for a large absorption-plus-scattering cross-section, as in the case made above for water clusters as Rydberg baryonic dark matter. Above 1.7 THz, the virtual photons of vacuum fluctuations would be gravitationally inactive because they are absorbed and scattered by the water clusters. Thus cosmic water buckyballs (Figs.1, 3, 4) and filamentary water clusters (Fig. 6) could be a common physical basis for both dark matter and dark energy. Because life as we know it is associated with "structured" or clustered water, this scenario is consistent with the "anthropic principle" that the Universe must have those properties which allow life to develop at some stage of its history [42].

**Figure Captions**

**Fig. 1.** Ground-state density-functional molecular-orbital states and vibrational modes of the $(H_2O)_{21}H^+$ protonated water "buckyball". **a.** Cluster molecular-orbital energy levels. The HOMO-LUMO energy gap is approximately 3eV. **b.** Wavefunctions of the lowest unoccupied cluster molecular orbitals. **c.** Vibrational spectrum. **d.** The 1.5 THz vibrational mode. The vectors show the directions and relative amplitudes of the O-O-O "bending" oscillations of the cluster "surface" oxygen atoms (in red) coupled to the hydronium ($H_3O^+$) oxygen ion vibration.

**Fig. 2.** **(a)** "Squashing" and "twisting" vibrational modes of a pentagonal dodecahedron. Hg and Hu designate the key irreducible representations of the icosahedral point group corresponding to these modes.
**(b)** Schematic representation of the potential energy wells for Jahn-Teller distorted water buckyballs and the resulting reduction of the energy barrier for chemical reaction of these water cluster along the reaction path defined by the normal mode coordinates $Q_s$.

**Fig. 3. a.** Pentagonal dodecahedral $(H_2O)_{20}$ "water buckyball". **b.** An array of three dodecahedral water clusters. **c.** An array of five dodecahedral water clusters.

**Fig. 4. a.** Vibrational spectrum of a pentagonal dodecahedral $(H_2O)_{20}$ "water buckyball". **b.** Lowest-frequency THz vibrational mode of the cluster. **c.** Vibrational spectrum of an array of three dodecahedral water clusters. **d**. Lowest-frequency THz vibrational mode. **e.** Vibrational spectrum of an array of five dodecahedral water clusters. **f.** Lowest-frequency THz vibrational mode. The vectors show the directions and relative amplitudes for the O-O-O "bending" motions responsible for the "squashing" mode of the cluster "surface" oxygen atoms.

**Fig. 5. a.** Pentagonal dodecahedral water buckyball cluster clathrating a methane molecule (carbon in blue), the common structural element of methane hydrate. **b.** Lowest-frequency THz vibrational mode of the methane clathrate. **c.** Water buckyball interacting with anthracene, a diesel combustion particulate matter ("soot") precursor. **d.** The vibrational coupling the water buckyball lowest-frequency THz "squashing" mode to a THZ "bending" mode of an anthracene soot precursor molecule. **e.** "Hemispherical" pentagonal dodecahedral water buckyball clathrating a "valyl" amino acid residue in insulin. **f.** Vibrational coupling of the water clathrate lowest-frequency THz "squashing" mode to a valyl THZ "bending" mode.

**Fig. 6. a.** Linear array of water clusters under periodic boundary conditions. **b.** Vibrational spectrum of the array with added electron to simulate electron donation from a surrounding microtubule. **c.** Lowest-frequency THz vibrational mode of the array to simulate electron donation from a surrounding microtubule.

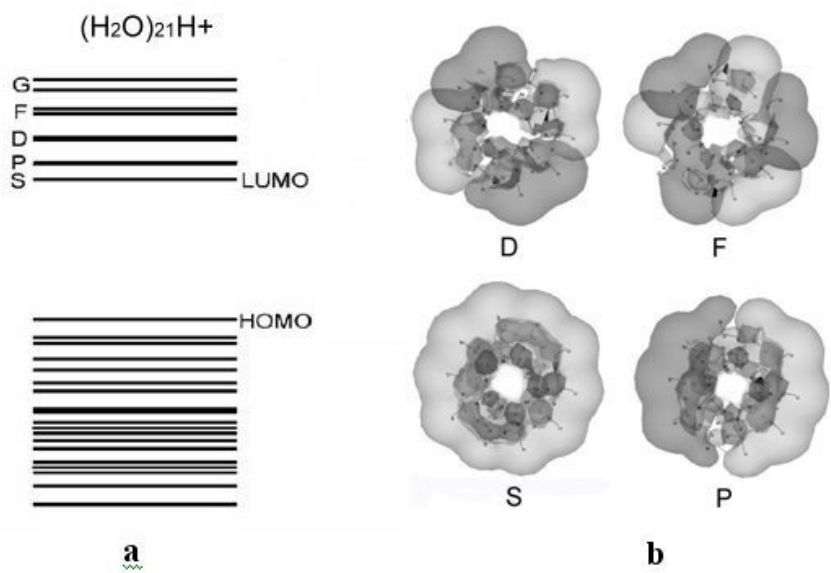
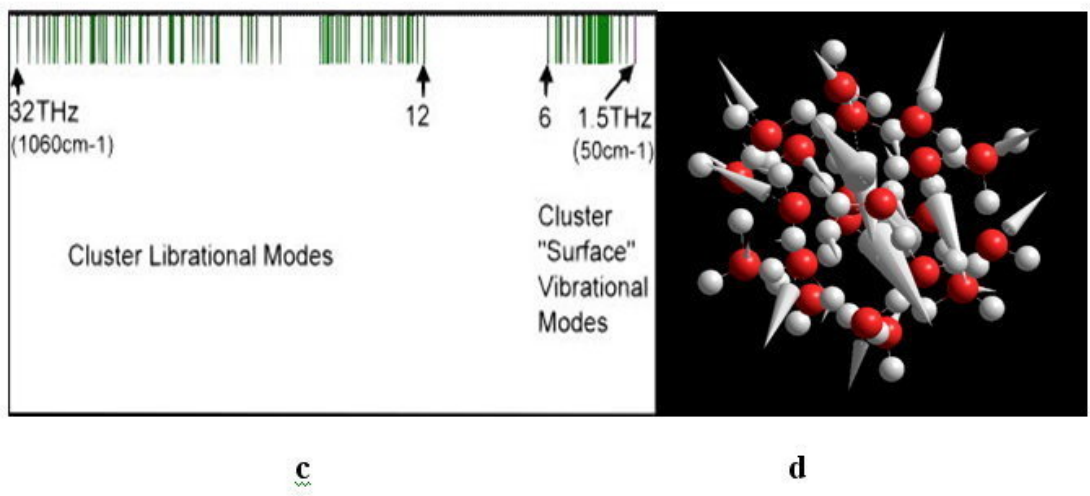

**Figure 1**

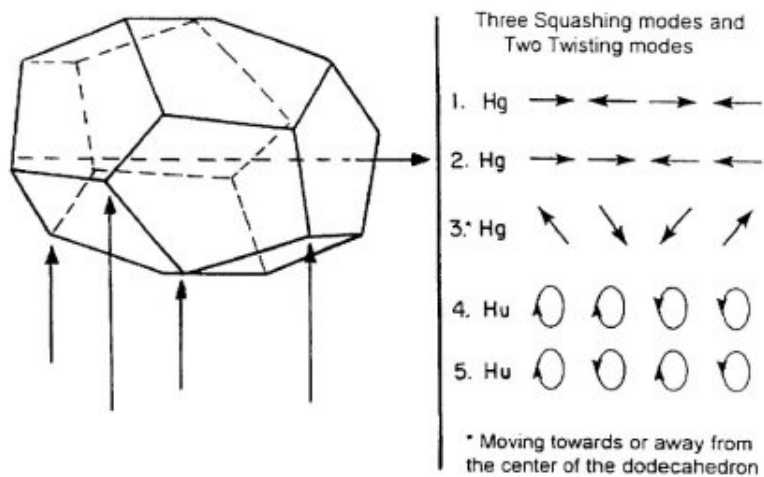

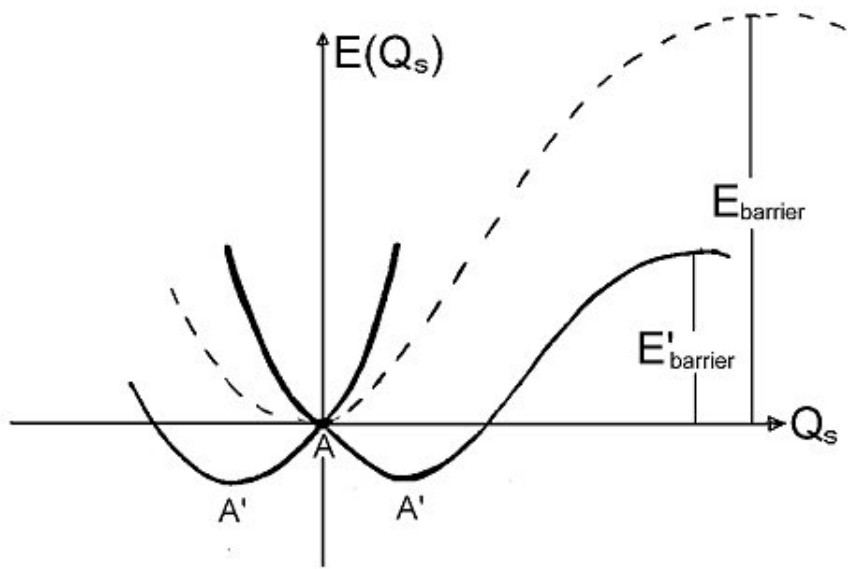

**Figure 2**

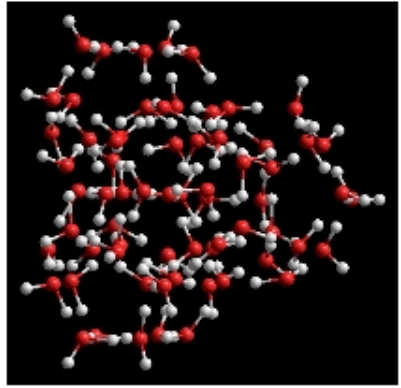

c

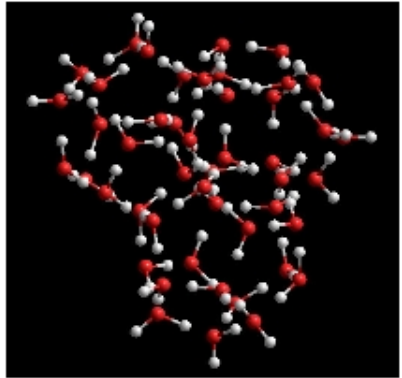

b

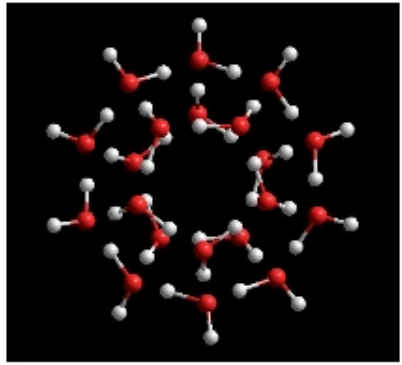

a

**Figure 3**

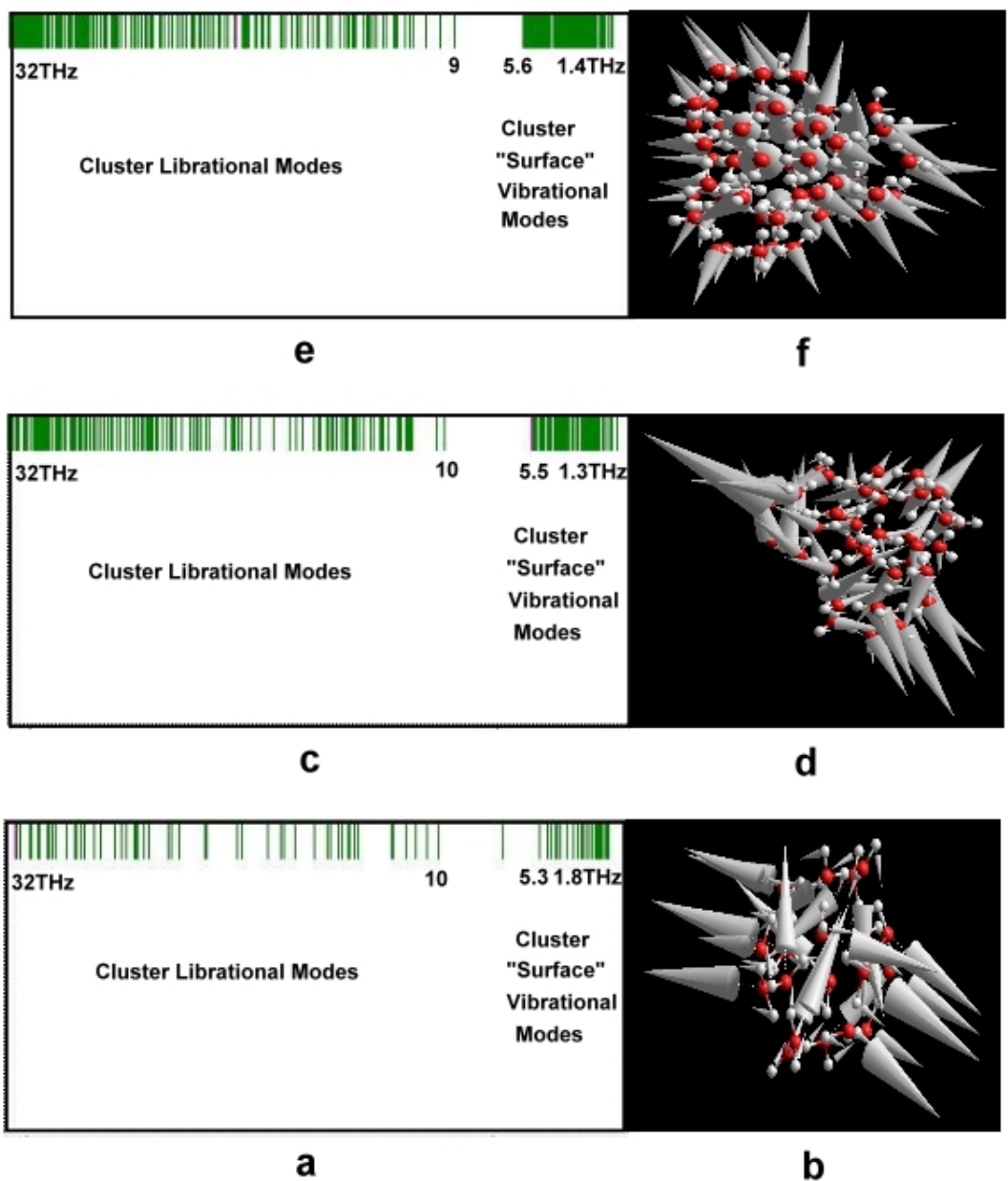

**Figure 4**

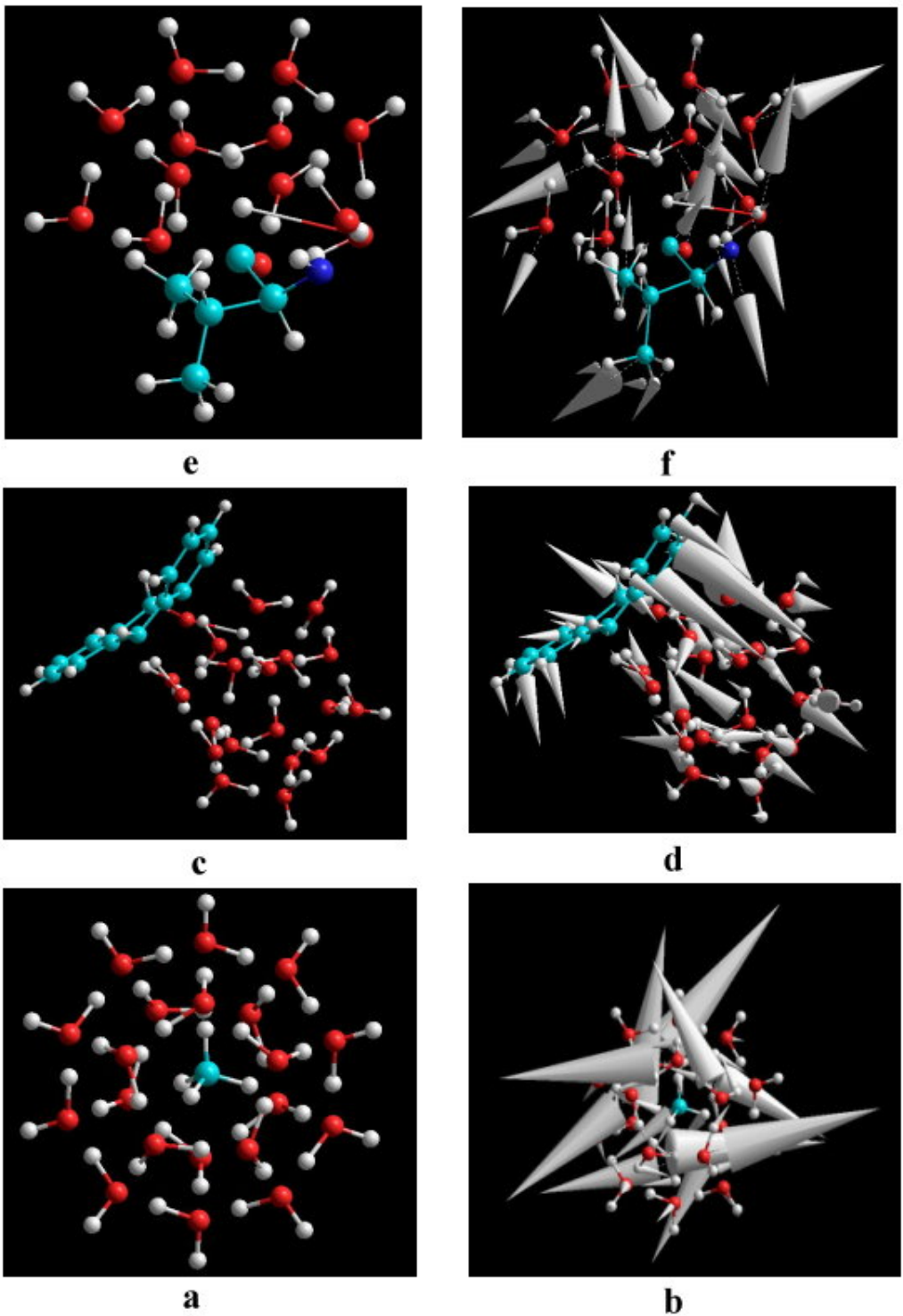

**Figure 5**

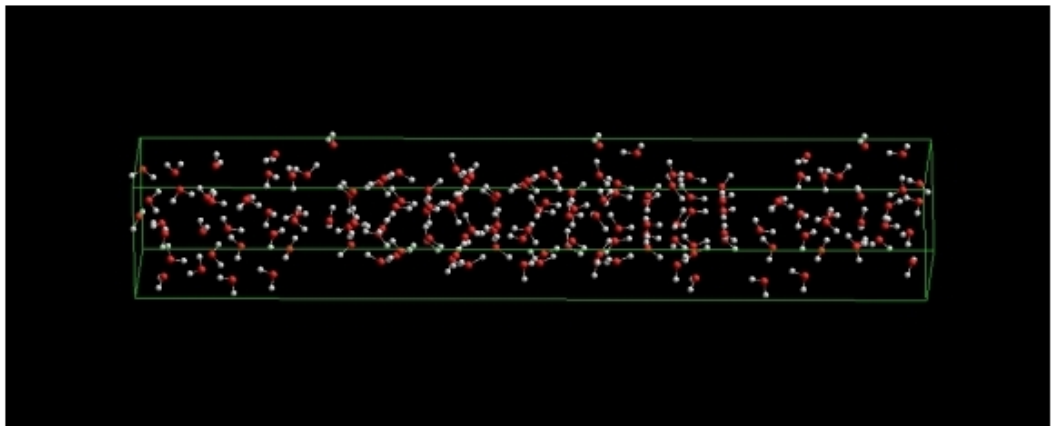
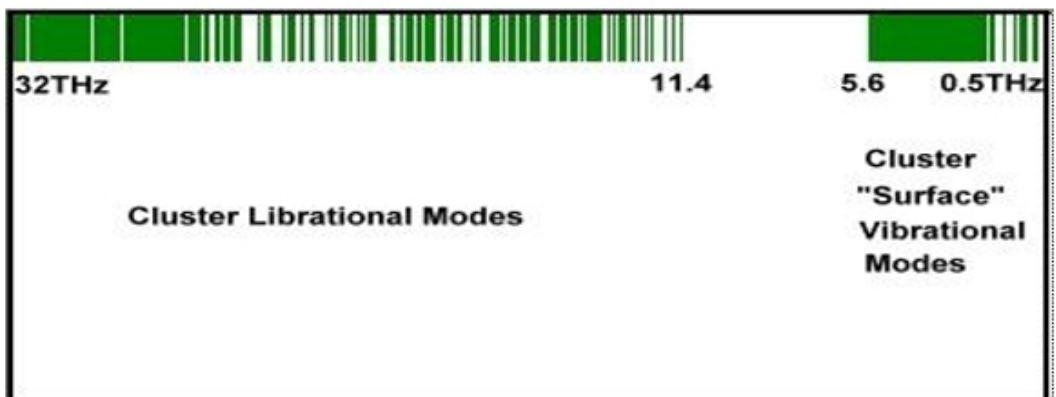
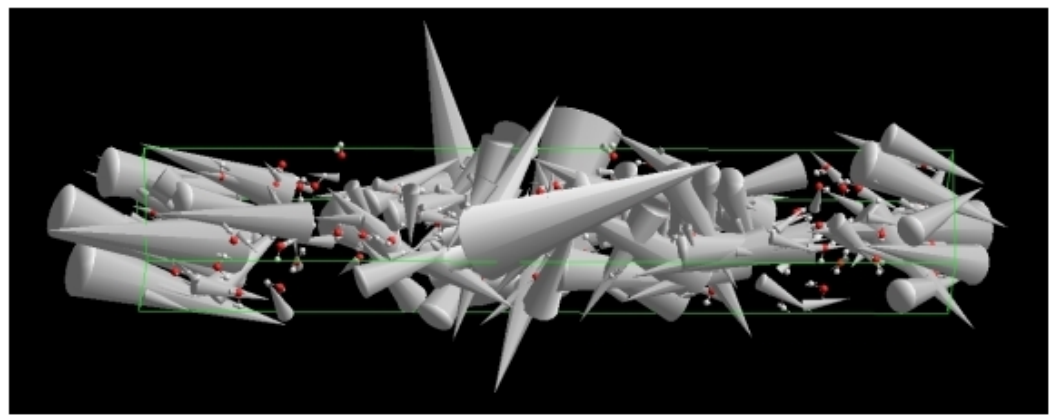

**Figure 6**